\documentclass[aps,showpacs,amssymb,eqsecnum,showkeys]{revtex4}
\usepackage{graphicx}
\usepackage[british]{babel}
\usepackage{amssymb}
\usepackage{amsmath}
\usepackage{amsfonts}
\usepackage[usenames]{color}
\usepackage{mathrsfs}

\begin{document}
\title{Patterns in the Kardar-Parisi-Zhang equation}
\author{Hans C. Fogedby}
\affiliation{Department of Physics and Astronomy, Aarhus University
\\
DK-8200, Aarhus C, Denmark\\
and
\\
Niels Bohr Institute, University of Copenhagen\\ DK-2100, Copenhagen
\O, Denmark}

\keywords{scaling, weak noise, growth modes, dynamical network,
upper critical dimension, non linear Schr\"odinger equation, domain
walls, solitons, dispersion, diffusive modes}

\pacs{05.40.-a, 02.50.-r, 05.45.Yv, 05.90.+m}

\begin{abstract}
We review a recent asymptotic weak noise approach to the
Kardar-Parisi-Zhang equation for the kinetic growth of an interface
in higher dimensions. The weak noise approach provides a many body
picture of a growing interface in terms of a network of localized
growth modes. Scaling in 1d is associated with a gapless domain wall
mode. The method also provides an independent argument for the
existence of an upper critical dimension.
\end{abstract}

\maketitle
\section{Introduction}
Non equilibrium phenomena are on the agenda in modern statistical
physics, soft condensed matter and biophysics. Open systems driven
far from equilibrium are ubiquitous. A classical case is driven
Navier Stokes turbulence, other cases are driven lattice gases,
growing interfaces, growing fractals, etc. Unlike equilibrium
physics where the Boltzmann-Gibbs scheme applies, the ensemble is
not known in non equilibrium. Here the problem is defined in terms
of a numerical algorithm, a master equation, or a Langevin
equation.

An interesting class of non equilibrium systems exhibit scale
invariance. One example is diffusion limited aggregation (DLA)
driven by the accretion of random walkers yielding a growing scale
invariant fractal with dimension $D\approx 1.7$ in 2d. Another
case is a growing interface driven by random deposition or
propagating in a random environment. Here the width of the growing
front $w(L,t)$ conforms to the dynamical scaling hypothesis
$w(L,t)=L^\zeta f(t/L^z)$, where $L$ is the size of the system and
$\zeta$ and $z$ scaling exponents; $\zeta$ characterizing the

roughness and $z$ describing the dynamical crossover to the
stationary profile \cite{inter}.

In the present paper we focus on the Kardar-Parisi- Zhang (KPZ)
equation which describes an intrinsic non equilibrium problem and
plays the same role as the Ginzburg-Landau functional in equilibrium
physics. The KPZ equation was introduced in 1986 in a seminal paper
by Kardar, Parisi and Zhang \cite{kpz}. It has the form
\begin{eqnarray}
\frac{\partial h}{\partial t} =
\nu\nabla^2h+\frac{\lambda}{2}\vec\nabla h\vec\nabla h -F+\eta,~~
\langle\eta\eta\rangle (\vec r,t)=\Delta\delta(\vec r)\delta(t),
\label{kpz}
\end{eqnarray}
and purports to describe non equilibrium aspects of a growing
interface; see Refs. \cite{inter,halpin}. Here $h(\vec r,t)$ is
the height of an interface at position $\vec r$ and time $t$, the
linear diffusion term $\nu\nabla^2h$, characterized by the
diffusion coefficient $\nu$, represents a surface tension, the
nonlinear growth term $(\lambda/2)\vec\nabla h \vec\nabla h$,
characterized by $\lambda$, is required to account for the lateral
growth, $F$ is an imposed constant drift, and the random aspects,
i.e., the random deposition of material or the random character of
the medium, are encoded in the noise $\eta(\vec r,t)$. The noise
is assumed to be locally correlated in space, and time, its
strength characterized by $\Delta$.

Despite its simple form the KPZ equation is difficult to analyze
and many aspects remain poorly understood \cite{inter,kpz,halpin}.
Apart from its intrinsic interest the KPZ equation is also related
to fundamental issues in {\em turbulence} and {\em disorder}.
Introducing the local slope field $\vec u=\vec\nabla h$ the KPZ
equation takes the form of a Burgers equation driven by conserved
noise,
\begin{eqnarray}
\frac{\partial\vec u}{\partial t}-\lambda(\vec
u\cdot\vec\nabla)\vec u = \nu\nabla^2\vec u+\vec\nabla\eta.
\label{burgers}
\end{eqnarray}
In the noiseless case setting $\lambda=-1$ and regarding $\vec u$
as a velocity field, the non linear term appears as the convective
term in the Navier-Stokes equation and Eq. (\ref{burgers}) has
been used to model aspects of {\em turbulence}. In the 1d case the
relaxation of the velocity field takes place subject to a
transient pattern formation composed of domain walls and ramps
with superimposed diffusive modes. In the driven case Eq.
(\ref{burgers}) was studied earlier in the context of long time
tails in hydrodynamics \cite{hydro}. On the other hand, applying
the non linear Cole-Hopf transformation the KPZ equation maps to
the Cole-Hopf equation (CH)
\begin{eqnarray}
\frac{\partial w}{\partial t} = \nu\nabla^2\vec
u-\frac{\lambda}{2\nu}wF+
\frac{\lambda}{2\nu}w\eta,~~h=\frac{2\nu}{\lambda}\ln w,
\label{cheq}
\end{eqnarray}
a linear diffusion equation driven by multiplicative noise. The CH
equation has a formal path integral solution \cite{kpz,halpin}
which can be interpreted as an equilibrium system of directed
polymers (DP) with line tension $1/4\nu$ in a quenched random
potential $\eta$; a model system in the theory of {\em disorder}
which has been studied using replica techniques
\cite{halpin,kardar}.

\begin{figure}[htbp]
\includegraphics[width=0.9\hsize]{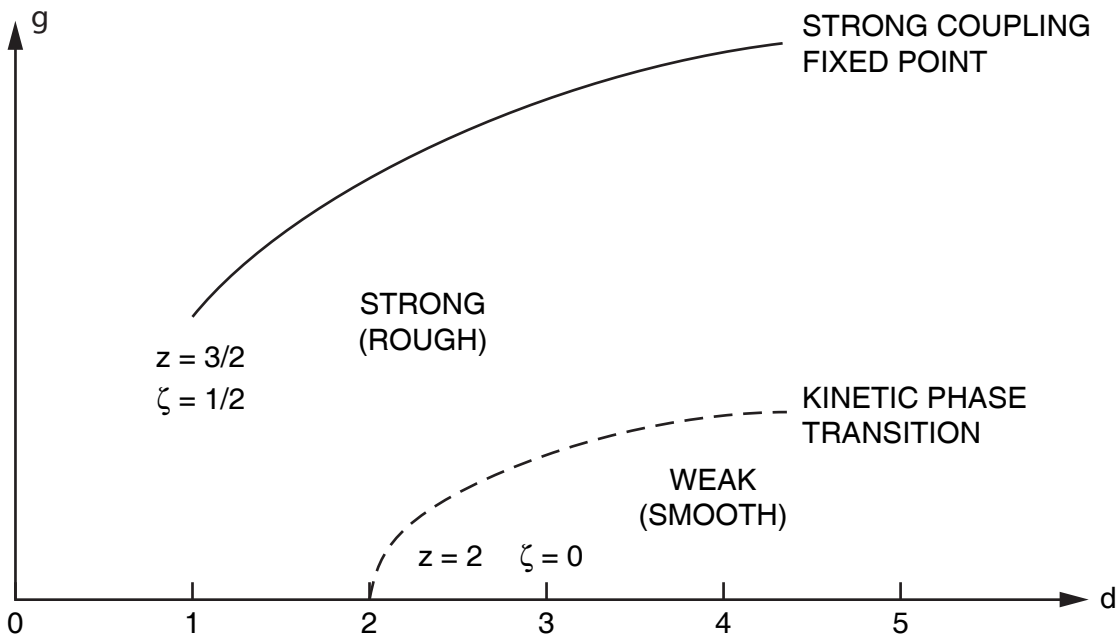}
\caption{DRG phase diagram for the KPZ equation to leading loop
order in $d-2$. In $d=1$ the DRG flow is towards the strong coupling
fixed point $\zeta=1/2, z=3/2$. Above the lower critical dimension
$d=2$ there is an unstable kinetic transition line, separating a
rough phase from a smooth phase.} \label{fig1}
\end{figure}

The KPZ equation lives at a critical point and conforms to the
dynamical scaling hypothesis. For the height correlations we have
\begin{eqnarray}
\langle hh\rangle(\vec r,t)=r^{2\zeta}F(t/r^z). \label{corr}
\end{eqnarray}
Here $\zeta$, $z$, and $F$ are the roughness exponent, dynamic
exponent, and scaling function, respectively. To extract scaling
properties the initial analysis of the KPZ equation was based on
the dynamic renormalization group (DRG) method, previously applied
to dynamical critical phenomena and noise-driven hydrodynamics
\cite{hydro}. An expansion in powers of $\lambda$ in combination
with a momentum shell integration yield to leading order in $d-2$
the DRG equation $dg/dl= \beta(g)$, with beta-function
$\beta(g)=(2-d)g+\text{const.}g^4$. Here $g=\Delta\lambda^2/\nu^3$
is the effective coupling strength and $l$ the logarithmic scale
parameter \cite{inter,kpz}. The emerging DRG phase diagram is
depicted in Fig. \ref{fig1}.

Before discussing the phase diagram we note two further properties
of the KPZ equation. First, subject to a Galilean transformation
the equation is invariant provided we add a constant slope to $h$
and adjust the drift $F$, i.e.,
\begin{eqnarray}
\vec r\rightarrow\vec r - \lambda\vec u^0t ,~~h\rightarrow h+ \vec
u^0\cdot\vec r,~~F\rightarrow F+(\lambda/2)(\vec u^0)^2;
\label{gal}
\end{eqnarray}
note that the slope field $\vec u$ and the diffusive field $w$
transform like $\vec u\rightarrow\vec u+\vec u^0$ and
$w\rightarrow w\exp[(\lambda/2\nu)\vec u^0\cdot\vec r]$,
respectively. The Galilean invariance implies the scaling law
\begin{eqnarray}
\zeta+z=2, \label{scal}
\end{eqnarray}
relating $\zeta$ and $z$ \cite{kpz}; the Galilean invariance is a
fundamental dynamical symmetry specific to the KPZ equation,
delimiting the universality class. Second, a
fluctuation-dissipation theorem is operational in 1d since the
stationary Fokker-Planck equation admits the explicit solution
\cite{kpz2}
\begin{eqnarray}
P_0(h)\propto\exp\left[-(\nu/\Delta)\int dx(\nabla h)^2\right].
\label{stat}
\end{eqnarray}
This distribution shows that the slope $u=\nabla h$ fluctuations
are uncorrelated and that the height field $h=\int^x udx'$
performs a random walk. Consequently, from Eq. (\ref{corr}) we
infer the roughness exponent $\zeta=1/2$ and from the scaling law
(\ref{scal}) the dynamic exponent $z=3/2$. In other words, the
scaling exponents associated with the strong coupling fixed point
are exactly known in 1d, see Fig. \ref{fig1}; moreover, results
for the scaling function can be obtained by loop expansions
\cite{kpz2}.

The lower critical dimension is $d=2$. Below $d=2$ the DRG flow is
towards the strong coupling fixed point. Above $d=2$ the DRG
equation yields an unstable kinetic phase transition line as
indicated in Fig. \ref{fig1}. For $\lambda$ below a critical
coupling strength $\lambda_c$ the DRG flow is towards $\lambda=0$,
corresponding to the linear Edwards-Wilkinson (EW) equation, the
KPZ equation for $\lambda =0$, yielding $z=2$ and $\zeta=(2-d)/2$,
note that the scaling law is not operational for $\lambda=0$; for
$\lambda>\lambda_c$ the DRG flow is towards a non perturbative
strong coupling fixed point.

A DRG analysis to all order in $d-2$ yields $z=2$ and $\zeta=0$ on
the transition line and a singularity at the upper critical
dimension $d_{\text{upper}}=4$. Mode coupling techniques also give
$d_{\text{upper}}=4$, whereas a directed polymer analysis yields
$d_{\text{upper}}\approx 2.5$ \cite{kpz2}.

In order to disentangle the properties of the KPZ equation, both
regarding scaling and otherwise, there is a need for alternative
methods. In the present paper we summarize a non perturbative weak
noise approach which we have pursued in recent years
\cite{Fogedby}. The working hypothesis here is to focus on the
noise strength $\Delta$ as the determining parameter rather that
the non linearity $\lambda$ and ensuing DRG analysis or mapping to
DP. For $\Delta=0$ the interface relaxes subject to a transient
pattern formation; for $\Delta\approx 0$ the interfaces initially
decays but is eventually driven into a stationary fluctuating
state; the crossover time diverging for $\Delta\rightarrow 0$. In
the next section we summarize the scheme that will allow us to
access the weak noise regime in a non perturbative fashion.

\section{Weak noise scheme}
\begin{figure}
\includegraphics[width=0.9\hsize]{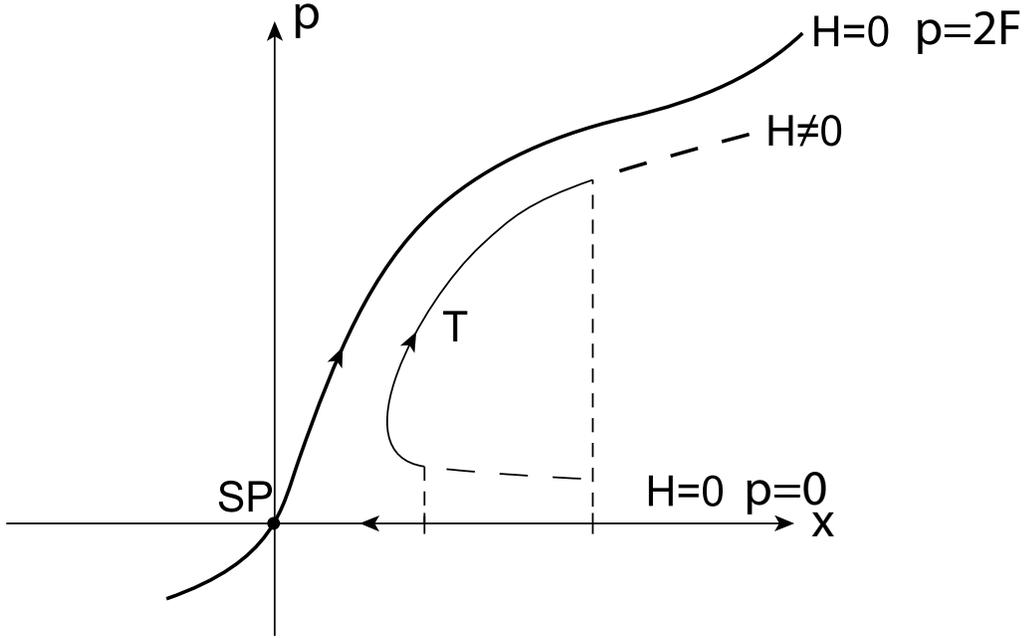}
\caption{Canonical phase space plot. The finite-time orbit from the
initial $x1$ to the final $x$ in transition time $T$ lies on the
energy manifold $H\neq 0$. In the long time limit the orbit migrates
to the zero-energy manifold composed of a transient submanifold for
$p=0$, corresponding to the noiseless case, and a stationary
manifold for $p=2F$, corresponding to the noisy case. The
submanifolds intersect in the saddle point (SP), determining the
Markovian behavior.} \label{fig2}
\end{figure}

The weak noise scheme takes as its starting point a generic Langevin
equation
\begin{eqnarray}
\frac{dx}{dt}=-F(x)+\eta,~~\langle\eta\eta\rangle
(t)=\Delta\delta(t), \label{lan}
\end{eqnarray}
determining the problem we wish to analyze. Here $x$ is a
multi-dimensional random variable, $F(x)$ a general non linear
drift, and $\eta$ additive white noise correlated with strength
$\Delta$. In order to implement a weak noise approximation we
consider the equivalent Fokker-Planck equation for the
distribution $P(x,t)$
\begin{eqnarray}
\Delta\frac{\partial P}{\partial
t}=\frac{1}{2}\Delta^2\frac{\partial^2P}{\partial
x^2}+\Delta\frac{\partial}{\partial x}(FP). \label{fp}
\end{eqnarray}
Interpreting $\Delta\partial/\partial x$ as a momentum operator, $P$
as an effective wave function, and $\Delta$ as an effective Planck
constant, Eq. (\ref{fp}) has the form of an imaginary time
Schr\"odinger equation and it is natural to introduce the well-known
WKB or eikonal approximation
\begin{eqnarray}
P\propto\exp[-S/\Delta]. \label{wkb}
\end{eqnarray}
To leading order in $\Delta$ the action then obeys a principle of
least action $\delta S=0$ as expressed by the Hamilton-Jacobi
equation $\partial S/\partial t+H(x,p)=0$ with canonical momentum
$p=\partial S/\partial x$. The Hamiltonian (energy) takes the the
form
\begin{eqnarray}
H=\frac{1}{2}p^2-pF(x)=\frac{1}{2}p[p-2F(x)], \label{ham}
\end{eqnarray}
yielding the coupled equations of motion
\begin{eqnarray}
\frac{dx}{dt}=-F+p,~~\frac{dp}{dt}=p\frac{dF}{dx}. \label{eqm}
\end{eqnarray}
The action associated with an orbit from $x_1$ to $x$ in time $T$
is
\begin{eqnarray}
S(x_1\rightarrow
x,T)=\int_{x_1,0}^{x,T}dt\left[p\frac{dx}{dt}-H\right]=
\frac{1}{2}\int_{x_1,0}^{x,T}dt p(t)^2. \label{act}
\end{eqnarray}
The weak noise recipe is clear. We solve the equations of motion
(\ref{eqm}) and identify an orbit from $x_1$ to $x$ in time $T$
with $p$ as an 'adjusted' variable. The orbits lie on constant $H$
manifolds. Evaluating the action $S$ for a specific orbit the WKB
approximation (\ref{wkb}) yields the transition probability from
$x_1$ to $x$ in time $T$. Assuming $F(x)\propto x$ for small $x$
the phase space has the generic structure depicted in Fig.
\ref{fig2}.

The present variationally based weak noise scheme, dates back to
Onsager. In more recent formulations it corresponds to the
saddlepoint contribution (optimal path) in the functional
Martin-Siggia-Rose scheme, see Refs. \cite{Fogedby,weaknoise}.

\section{Growth modes}
The weak noise scheme applies directly to the KPZ equation in the
Cole-Hopf formulation (\ref{cheq}). Extending the scheme in order
to incorporate multiplicative noise, see Refs. \cite{Fogedby}, the
WKB scheme yields the equations of motion, action, and
distribution
\begin{eqnarray}
&&\frac{\partial w}{\partial t} = \nu[\nabla^2w-k^2w]+k_0^2w^2p,~~
\frac{\partial p}{\partial t}=-\nu[\nabla^2 w-k^2w]-k_0^2pw^2,
\label{eqmo}
\\
&& S(w,T)=(k_0^2/2)\int^{w,T}d \vec r
dt(wp)^2,~~P(w,T)\propto\exp[-S(w,T)/\Delta],
\end{eqnarray}
with parameters $k^2=\lambda F/2\nu$ and $k_0=\lambda/2\nu$. On
the transient and stationary manifolds $p=0$ and $p\propto w$ the
equations of motion reduce in the static case to the diffusion and
non linear Schr\"odinger equations
\begin{eqnarray}
&&\nabla^2w=k^2w, \label{diff}
\\
&&\nabla^2w=k^2w-k_0^2w^3, \label{nse}
\end{eqnarray}
admitting localized spherically symmetric solutions
$w_+(r)\propto\exp(kr)$ and $w_-(r)\propto\exp(-kr)$ for large
$r$, respectively. In terms of the height field $h=k_0^{-1}\ln w$
and slope field $\vec u=\vec\nabla h$ we obtain the fundamental
localized static growth modes for large $r$
\begin{eqnarray}
h_\pm(r)\propto\pm\frac{k}{k_0}r~~ \text{and} ~~\vec
u_\pm(r)\propto\pm\frac{k}{k_0}\frac{\vec r}{r}.
\end{eqnarray}
The growth modes are characterized by the amplitude or charge
$k\propto F^{1/2}$, determined by the imposed drift $F$ in the KPZ
equation. The growth mode with positive charge
$h\propto\exp(kr/k_0)$, $\vec u\propto \vec r/r$ lives on the
transient 'noiseless' $p=0$ manifold and corresponds to a cone or
dip in $h$ and a constant positive monopole in $\vec u$. The mode
carries zero action, zero energy $H=-S/T$=0, and zero momentum
$\vec\Pi=\int d\vec rw\vec\nabla p=0$; the distribution
$P\propto\exp(-S/\Delta)$ associated with the mode is of $O(1)$.
The growth mode with negative charge $h\propto\exp(-kr/k_0)$,
$\vec u\propto -\vec r/r$ is associated with the stationary
'noisy' manifold $p\propto w$ and corresponds to an inverted cone
or tip in $h$ and a constant negative monopole in $\vec u$. This
modes carries a finite action
\begin{eqnarray}
S\propto T(\nu/k_0)^2k^{4-d}, \label{actgm}
\end{eqnarray}
yielding the distribution $P\propto\exp[-S/\Delta]$.

\begin{figure}
\includegraphics[width=0.9\hsize]{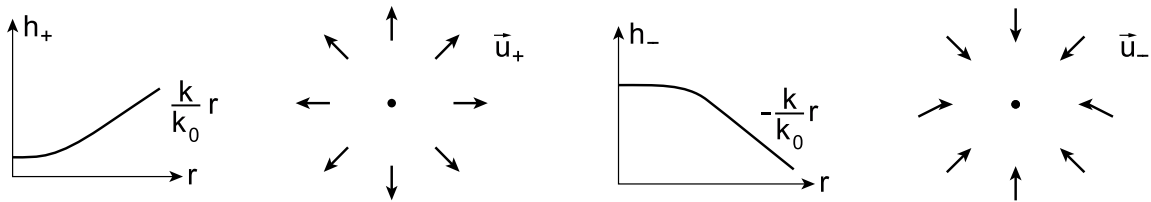}
\caption{We depict the static growth modes in the height and slope
fields in 2d.} \label{fig3}
\end{figure}

In the 1d case the growth modes correspond to right hand and left
hand domain walls, $u\propto\pm\tanh(kx)$. The right hand domain
wall is the viscosity broadened shock wave in the deterministic
Burgers equation; the left hand domain wall is 'noise induced', in
the present scheme characterized by a finite $p$, see Refs.
\cite{Fogedby}. In Fig. \ref{fig3} we have depicted the static
growth modes in the height and slope fields.

Applying the Galilean transformation (\ref{gal}) we can boost the
static modes and obtain the propagating growth modes
\begin{eqnarray}
h_\pm(r,t)=\pm\frac{k}{k_0}|\vec r+\lambda\vec u_0t|+\vec
u_0\cdot\vec r,~~ u_\pm(r,t)= \pm\frac{k}{k_0}\frac{\vec
r+\lambda\vec u_0t}{|\vec r+\lambda\vec u_0t|}+\vec u_0.
\label{gm}
\end{eqnarray}
The moving localized growth modes are the fundamental elementary
excitations incorporating the non linear aspects of the KPZ
equation. In a quantum field theory context the growth modes
correspond to instantons or solitons.

\section{Stochastic pattern formation}
By means of the propagating localized growth modes (\ref{gm}) we
construct a global solution of the field equations (\ref{eqmo}). The
Galilean invariance (\ref{gal}) determines the matching of the modes
and we obtain the dilute network solution
\begin{eqnarray}
&&h(\vec r,t)=k_0^{-1}\sum_ik_i|\vec r-\vec r_i(t)|, ~~ \vec u
(\vec r,t)=k_0^{-1}\sum_i k_i\frac{\vec r-\vec r_i(t)}{|\vec
r-\vec r_i(t)|}, \label{net1}
\\
&&\vec v_i(t)=-2\nu\sum_{l\neq i}k_l\frac{\vec r_i(t)-\vec
r_l(t)}{|\vec r_i(t)-\vec r_l(t)|},~~\vec r_i(t)=\int_0^t\vec
v_i(t')dt'+\vec r_i(0). \label{net2}
\end{eqnarray}
Here $k_i$ is the assignment of charges and $\vec r_i(0)$ the
initial positions. This network construction correspond to a
multi-instanton solution in quantum field theory. As time evolves
the modes propagate and the velocities adjust to constant values
given by the self-consistent equation
\begin{eqnarray}
\vec v_i=-2\nu\sum_{l\neq i}k_l\frac{\vec v_i-\vec v_l}{|\vec
v_i-\vec v_l|}.
\end{eqnarray}
Superimposed on the propagating network of growth modes is a
spectrum of linear extended diffusive modes with dispersion
$\omega=\nu k^2$, following from a linear analysis of the field
equations (\ref{eqmo}).

For large $\vec r$ the slope field  $\vec u\sim (\vec
r)/r)\sum_ik_i$. In order to ensure the boundary condition of a
flat interface at large distances we impose the neutrality
condition $\sum_ik_i=0$; note that this boundary condition still
allows for a local offset of $h$ and the propagation of facets or
steps. During time evolution the dynamical network propagates
across the system. Imposing periodic or bouncing boundary
conditions increments are added to $h$ and the interface grows,
see also Ref. \cite{Fogedby}.

From an analysis of Eqs. (\ref{net1}) and (\ref{net2}) it follows
that the growth modes with positive charge exerts an attraction of
the other modes, whereas the negatively charged growth modes repel
the other modes.  Although our analysis so far only applies to a
dilute network, it follows tentatively from a numerical
simulations of Eqs. (\ref{net1}) and (\ref{net2}) that the modes
form dipoles and that the long time stable network configuration
is composed of a gas of propagating dipoles, i.e. the pairing of
monopole growth modes with opposite charges.

\begin{figure}
\includegraphics[width=0.9\hsize]{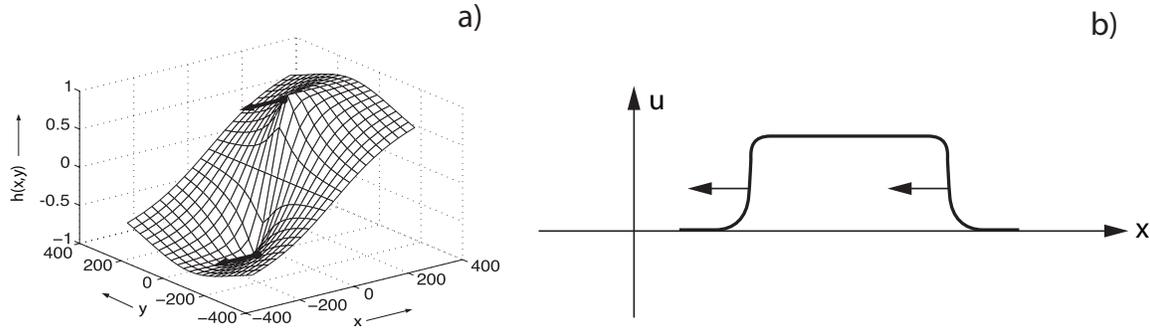}
\caption{We show a dipole configuration composed of two paired
monopoles of opposite charges. In a) we depict the dipole mode for
the height field in 2d; in b) we show the two-domain wall
configuration for the slope field in 1d.} \label{fig4}
\end{figure}

In terms of the slope field a dipole mode with charges $k$ and
$-k$ has the form $u_{\text{dip}}\sim (k/k_0)[(\vec r-\vec vt-\vec
r_1)/|\vec r-\vec vt-\vec r_1|-(\vec r-\vec vt-\vec r_2)/|\vec
r-\vec vt-\vec r_2|]$, $\vec v=\lambda(k/k_0)(\vec r_2-\vec
r_1)/|\vec r_2-\vec r_1|$. The mode propagate with velocity $v$
and carries energy, momentum and action,
$E_{\text{dip}}\propto(\nu^2/k_0^2)k^{4-d}$,
$\Pi_{\text{dip}}\propto(\nu/k_0^2)k^{3-d}$, and
$S_{\text{dip}}\propto(\nu^2/k_0^2)k^{4-d}T$. In terms of the
height field $h_{\text{dip}}\sim (k/k_0)[|\vec r-\vec vt-\vec
r_1|-|\vec r-\vec vt-\vec r_2|]$ the mode in depicted in Fig.
\ref{fig4}. Asymptotically the height field is flat corresponding
to a vanishing slope field. The dipole mode corresponds to a
propagating local defect or deformation. In 1d the dipole mode
corresponds in the slope field to a matched right hand and left
hand domain wall propagating across the system, see Fig.
\ref{fig4}.

\section{Scaling and upper critical dimension}

The scaling issues for the KPZ equation remain unsettled except in
1d where the fluctuation-dissipation theorem (\ref{stat}) and the
scaling law (\ref{scal}) give access to the strong coupling fixed
point with scaling exponents $\zeta=1/2$ and $=3/2$, see Fig.
\ref{fig1}. In higher d there has been many attempts to access the
strong coupling fixed point both on the basis of DRG, mode
coupling, directed polymers (DP), and numerically; however, the
strong coupling features remain elusive \cite{kpz2}.

The present weak noise method is not a scaling approach but rather
a many body description of a growing interface. Nevertheless, the
method allows a discussion of some of the scaling features. Since
the scheme is consistently Galilean invariant the scaling law
(\ref{scal}) is automatically obeyed. The roughness exponent
$\zeta$ is associated with the static correlations $\langle
hh\rangle(\vec r)=\int\prod dh h(\vec r)h(0)P_0(h)\propto
r^{\zeta}$ and requires the static distribution $P_0(h)$,
$P_0(h)\propto\exp[-S_0(h)/\Delta]$,
$S_0(h)=\lim_{T\rightarrow\infty}S(h_1\rightarrow h,T)$. The
stationary distribution is associated with the zero-energy
stationary manifold which in general is difficult to identify for
a system with many degrees of freedom.

In 1d in terms of the slope field $u=\nabla h$ and associated
noise field $p$ the Hamiltonian takes the form $H=\int
dx~p[\nu\nabla^2u+\lambda u\nabla u-(1/2)\nabla^2p]$ and it
follows that the stationary zero-energy manifold is given by
$p=2\nu u$, yielding a Hamiltonian density as a total
differential. Correspondingly, the stationary action
$S_0\propto\int dx dtp\partial u/\partial t = \nu\int dx(\nabla
h)^2$ and the distribution $P_0(u)\propto\exp[-S_0/\Delta]$ in
accordance with (\ref{stat}), implying $\zeta=1/2$. Since Galilean
invariance is built in the exponent $z=3/2$ follows automatically.
However, we can also infer $z=3/2$ from an independent argument
based on the dispersion law for the low lying gapless excitations.
The elementary excitations are the right hand and left hand domain
walls. A composite quasi particle or dipole mode satisfying the
boundary condition of vanishing slope can be constructed by paring
two domain walls. The dipole mode propagates with energy $E\propto
u^3$ and momentum $\Pi\propto u^2$, where $u$ is the dipole
amplitude. Using the analogy between the stochastic formulation
and quantum mechanics, i.e., the canonical quantization of the
weak noise scheme with $\Delta$ as an effective Planck constant,
and using the spectral representation $\langle uu\rangle(x,t)=\int
d\Pi F(\Pi)\exp(-Et/\Delta+i\Pi x/\Delta)$, $F(\Pi)$ is a form
factor, we note that the bottom of the gapless dipole energy
spectrum $E\propto\Pi^{3/2}$ implies $\langle
uu\rangle(x,t)\propto G(t/x^z)$, where $z=3/2$.

In higher $d$ we have not been able to identify the stationary
zero-energy manifold and the exponent $\zeta$. Also, in order to
determine $z$, for example from the form of the time dependent
correlations $\langle hh\rangle(\vec r,t)$, we need both the
transition probabilities and the stationary distribution, i.e., by
definition $\langle hh\rangle(\vec r,t)=\int\prod
dh_1dh_2P(h_1(0)\rightarrow h_2(\vec r),t)P_0(h_1)$. In the weak
noise scheme $P(h_1\rightarrow h_2,t)\propto\exp[-S(h_1\rightarrow
h_2,t)/\Delta]$ which requires a detailed analysis of the
dynamical network and the associated action.

In the dipole sector we can, however, present some preliminary
scaling results. Since the propagating dipole mode according to
(\ref{actgm}) carries action $S\propto k^{4-d}T$ and propagates
with velocity $v\propto k$ the mode moves the distance $L=vT$ in
time $T$. Expressing $S$ in the form $S\propto L^{4-d}/T^{3-d}$ we
infer the single dipole distribution $P\propto\exp[-\text{const}.
L^{4-d}/T^{3-d}]$ and, correspondingly, the dipole mean square
displacement $\langle\delta L^2\rangle\propto T^{2H}$ with Hurst
exponent $H=(3-d)/(4-d)$; note that the dynamic exponent
$z=H^{-1}=(4-d)/(3-d)$. In the stochastic representation the
dipole mode thus performs anomalous diffusion. In $d=0$ we have
$H=3/4$, $z=4/3$, in agreement with a formal DP result
\cite{halpin,kpz2}. In $d=1$ we obtain $H=2/3$ and the exact
result $z=3/2$, i.e., the dipole modes exhaust the spectrum. In
$d=2$ we have $H=1/2$ and $z=2$, i.e., ordinary diffusion. In
$d=3$ we obtain $H=0$ and $z=\infty$, the dipole mean square
displacement falls off logarithmically $\langle\delta
L^2\rangle\propto \ln T$, however, $z$ diverges at variance with
accepted DRG and DP results. In $d=4$ we have $H=-\infty$ and the
mean square displacement is arrested. Below $d=2$ (the lower
critical dimension) the dipole modes superdiffuse, above $d=2$ we
have subdiffusion. These scaling results only refer to the dipole
sector.

The last issue is the much discussed upper critical dimension for
the KPZ equation \cite{kpz2}. The weak noise approach allows a non
scaling argument for the existence of a critical dimension. Above
$d=4$ the negative growth mode as a bound state solution to the
non linear Schr\"odinger equation (\ref{nse}) ceases to exist.
This implies that the dynamical network representation of a
growing interface ceases to be valid. This result follows from a
numerical analysis of the non linear Schr\"odinger equations
(\ref{nse}) but can also be inferred by an algebraic proof based
on Derrick's theorem. First, introducing $K=(1/2)\int d^dx(\nabla
w)^2$, $N=\int d^dxw^2$, and $I=\int d^dxw^4$ we infer from
(\ref{nse}) the identity $-2K=k^2N-k_0^2I$. Second, deducing
(\ref{nse}) from a variational principle $\delta F/\delta w=0$
with $F=K+(1/2)k^2N-(k_0^2/4)I$ and performing a constrained
minimization $w(\vec r)\rightarrow w(\mu\vec r)$,
$K\rightarrow\mu^{d-2}K$, $N\rightarrow\mu^dN$, $I\rightarrow\mu^d
I$ and $\delta F/\delta \mu|_{\mu = 1}=0$ we have the second
identity $(d-2)K+(k^2/2)dN-(k_o^2/4)dI=0$. Finally, requiring
$N,I>0$ the identities imply $d<4$, q.e.d.

\section{Summary and conclusion}
In this paper we have presented a short review of a recently
developed asymptotic weak noise approach to the
Kardar-Parisi-Zhang equation. The scheme provides a many body
description of a growing interface in terms of a dynamical network
of growth modes. The growth modes are the elementary building
blocks and their propagation accounts for the kinetic growth.
Kinetic transitions are determined by an associated dynamical
action, replacing the customary free energy landscape.
Superimposed on the network is a gas of diffusive modes. In 1d the
dispersion laws delimit the universality classes: In the KPZ case
the gapless domain wall modes yield $z=3/2$, the diffusive modes
being subdominant; in the EW case the domain walls are absent and
the gapless diffusive modes yield $z=2$. In higher d the scaling
results based on the weak noise method are still subject to
scrutiny. Finally, we mention that the weak noise method has also
been applied to the noise-driven Ginzburg-Landau equation, a
finite-time-singularity model, and DNA bubble dynamics
\cite{fogedby2}.


\begin{thebibliography}{99}
\bibitem{inter} A. -L. Barabasi and H. E. Stanley,
{\em Fractal Concepts in Surface Growth} (Cambridge University
Press, 1995); J. Krug and H. Spohn, {\em Solids Far from
Equilibrium; Kinetic roughening of growing surfaces: Fractal
Concepts in Surface Growth} (Cambridge University Press, 1992); J.
Krug, Adv. Phys. {\bf 46}, 139 (1997)
\bibitem{kpz}
M. Kardar, G. Parisi and Y. C. Zhang, Phys. Rev. Lett. {\bf 56},
889 (1986); E. Medina, T. Hwa, M. Kardar and Y. C. Zhang, Phys.
Rev. A {\bf 39}, 3053 (1989)
\bibitem{halpin}
T. Halpin-Healy and Y. C. Zhang, Phys. Rep. {\bf 254}, 215 (1995)
\bibitem{hydro} D. Forster, D. R. Nelson and M. J. Stephen,
Phys. Rev. Lett. {\bf 36}, 867 (1976); Phys. Rev. A {bf 16}, 732
(1977)
\bibitem{kardar} M. Kardar and Y. C. Zhang, Phys. Rev. Lett. {\bf
58}, 2087 (1987); M. Kardar, Nucl. Phys. B {bf 290}, 582 (1987)
\bibitem{Fogedby}
H. C. Fogedby, Phys. Rev. Lett. {\bf 94},195702 (2005); Phys. Rev.
E {\bf 73}, 031104 (2006); Phys. Rev. E {\bf 68}, 026132 (2003);
Phys. Rev. E59, 5065 (1999); Phys. Rev. E57, 49431 (1998); Phys.
Rev. Lett. 80, 1126 (1998)
\bibitem{weaknoise}L. Onsager and S. Machlup,
Phys. Rev. {\bf 91}, 1505 (1953), ibid 1512; P C. Martin, E. D.
Siggia and H. A. Rose, Phys. Rev. A {\bf 8}, 423 (1973); R. Baussch,
H. K. Janssen and H. Wagner, Z. Phys. B {\bf 24}, 113 (1976)
\bibitem{kpz2} D. A. Huse, C. L. Henley and D. S. Fisher, Phys.
Rev. Lett. {\bf 55}, 2924 (1985); E. Frey and U. C. T\"{a}uber,
Phys. Rev. E {\bf 50}, 1024 (1994); E. Frey, U. C. T\"{a}uber and
T. Hwa, Phys. Rev. E {\bf 53}, 4424 (1996); K. J. Wiese, J. Stat.
Phys. {\bf 93}, 143 (1998); F. Colaiori and M. A. Moore, Phys.
Rev. Lett. {\bf 86}, 3946 (2001); M. L\"{a}ssig, Phys. Rev. Lett.
{\bf 80}, 2366 (1998); Nucl. Phys. B {\bf 448}, 559 (1995); M.
L\"{a}ssig and H. Kinzelbach, Phys. Rev. Lett. {\bf 78}, 903
(1997); P. Le Doussal and K.J.\ Wiese, Phys. Rev. E {\bf 72},
035101 (2005)
\bibitem{fogedby2} H. C. Fogedby, J. Hertz and A. Svane, Europhys.
Lett. {\bf 62}, 795 (2003); H. C. Fogedby and V. Poutkaradze, Phys.
Rev. E {\bf 66}, 021103 (2002); Hans C. Fogedby and Ralf Metzler,
Phys. Rev. Lett. {\bf 98}, 070601 (2007).
\end{thebibliography}
\end{document}